# PANDAVA:
# Semantic and Reflexive Protocol for Interdisciplinary and Cognitive Knowledge Synthesis

Eldar Knar[1]

Tengrion, Astana, Republic of Kazakhstan
eldarknar@gmail.com
https://orcid.org/0000-0002-7490-8375

**Abstract**

Modern science faces the need to move from linear systematic review protocols to deeper cognitive navigation across fields of knowledge. In this context, the PANDAVA protocol (Protocol for Analysis and Navigation of Deep Argumentative and Valued Knowledge) is designed for analysing the semantic structures of scientific knowledge. It combines semantic mapping, assessment of concept maturity, clustering, and generation of new hypotheses.

PANDAVA is interpreted as the first interdisciplinary protocol for knowledge systematization focused on semantic and cognitive mapping.

The PANDAVA protocol integrates quantitative analysis methods with reflective procedures for comprehending the structure of knowledge and is applied in interdisciplinary, theoretically saturated fields where traditional models such as PRISMA prove insufficient.

As an example, the protocol was applied to analyse the abiogenesis hypotheses. Modelling demonstrated how to structure theories of the origin of life through the integration of data on microlight, turbulent processes, and geochemical sources.

PANDAVA enables researchers to identify strong and weak concepts, construct knowledge maps, and develop new hypotheses.

Overall, PANDAVA represents a cognitively enriched tool for meaningful knowledge management, fostering the transition from the representation of facts to the design of new scientific paradigms.

**Declarations and Statements:**
No conflicts of interest
This work was not funded
No competing or financial interests
All the data used in this work are in the public domain.
Generative AI (LLM or other) was not used in writing the article. Except for the search engine *SciSpace* for Reference and the *Jupiter Notebook* environment for running our *Python* scenarios.
Ethics committee approval is not needed (without human or animal participation).

---

[1] PhD in Physics, Fellow of the Royal Asiatic Society of Great Britain and Ireland, Member of the Philosophy of Science Association (Baltimore, Maryland, USA)

## 1. Introduction

Modern science increasingly resembles an exponentially growing archipelago of knowledge, where paradoxically, both the volume of information (Bornmann, 2015) and epistemic fragmentation (Frickel & Gross, 2005) are simultaneously increasing. In a hyperproductive scientific environment, intensified by digital platforms, systematic reviews have become a key tool for the navigation, synthesis, and transformation of knowledge. However, existing systematic review protocols, such as PRISMA (Moher, 2009), GRADE (Guyatt, 2008), STROBE (von Elm, 2007), and their analogues, are oriented primarily toward the aggregation of empirical data, leaving aside the conceptual, ontological, and theoretical structure of the studies themselves.

Moreover, classical systematic review protocols typically assume a linear model of analysis (Gough et al., 2012), moving from a formalized query to source selection and their quantitative integration. This approach has proven extremely effective in medicine and biological sciences but is less applicable in contexts of complex, interdisciplinary, or theoretically saturated domains, such as philosophy, cognitive sciences, artificial intelligence, sociotechnical studies, and futuristics.

In response to this challenge, we propose a new protocol, PANDAVA (Protocol for Analysis and Navigation of Deep Argumentative and Valued Knowledge). This protocol is focused on semantic synthesis, ontological mapping, epistemic clustering, and the generation of new hypotheses.

PANDAVA represents a cognitively enriched platform that enables researchers not only to systematize the literature but also to comprehend its internal conceptual tensions, identify semantic gaps, and model potential trajectories of scientific synthesis.

PANDAVA relies on four key principles

*Semantic completeness, where not only publications but also their conceptual cores—hypotheses, models, ontologies, arguments—are analysed;*
*Epistemic navigation, which uses clustering by type of knowledge rather than keywords or methods;*
*Ontological reflection, creating a map of meanings, levels of analysis, and hidden assumptions;*
*Hypothesis generation makes the formulation of new research directions the ultimate goal rather than merely describing existing directions.*

In this article, we present the full architecture of the protocol as a tool for metascientific integration and meaningful thinking under conditions of information overload.

Essentially, the protocol stands independently. However, for clarity and better understanding, we first briefly consider PANDAVA as a cognitive extension of PRISMA.

Preferred Reporting Items for Systematic Reviews and Meta-Analyses (PRISMA) is an international standard for systematic reviews that focuses on the collection of empirical data, the rigorous selection of publications on the basis of formal criteria, a transparent inclusion/exclusion scheme, and the minimization of bias. The purpose of PRISMA is to make systematic reviews representative, verifiable, and evidence-based.

Let us highlight some limitations of PRISMA in the conceptual and theoretical domains (Table 1). In particular, although PRISMA is excellent for reviewing empirical studies, it has natural limitations.

**Table 1.** Natural limitations of PRISMA

| PRISMA Limitation | Consequence |
|---|---|
| Focus on quantitative research | Theoretical models remain outside the review |
| No analysis of concept maturity | Weak or pseudotheoretical ideas may be included |
| No semantic structure mapping | Loss of holistic understanding of the research field |
| No identification of semantic gaps | Inability to formulate new hypotheses |
| No visualization of cognitive landscapes | Loss of architectural thinking by the researcher |

Thus, in our view, PANDAVA does not compete with PRISMA but naturally extends its capabilities where needed (Table 2).

**Table 2.** PANDAVA in the context of the PRISMA extension

| PANDAVA Component | What it adds to PRISMA |
|---|---|
| P1: Problem Cartography | Structures not only articles but knowledge itself |
| A1: Argument Harvesting | Extracts theoretical constructs, not just data |
| N1: Navigation of Structures | Creates a network of conceptual interrelations instead of a flat list |
| D1: Deep Clustering | Clusters concepts by maturity and semantic strength |
| A2: Gap Mapping (Booth, 2015) | Identifies semantic gaps for new research |
| V1: Value-Oriented Synthesis | Constructs a holistic intellectual architecture of the field |
| A3: Hypothesis Generation | Generates new hypotheses based on gaps and synthesis |

In the context of complementarity, the maximum effect is achieved when PRISMA and PANDAVA are used together (Table 3).

**Table 3.** Effects of combining PRISMA + PANDAVA

| If using only PRISMA | If supplemented with PANDAVA |
|---|---|
| You obtain a list of reliable articles | You obtain a map of knowledge and new hypotheses |
| You minimize inclusion errors | You design a new theoretical space |
| You standardize the review | You enhance the review through reflection and synthesis |
| You reproduce existing knowledge | You contribute to the development of knowledge |

The conceptual distinction between PANDAVA and PRISMA can be interpreted as follows (Table 4):

**Table 4.** Semantic differentiation between PANDAVA and PRISMA

| Aspect | PRISMA | PANDAVA |
|---|---|---|
| Main metaphor | Selection of publications | Navigation through semantic structures (Börner, 2003) |
| Epistemological role | Representation of existing knowledge | Reconstruction and design of knowledge |
| Mode of thinking | Evaluation of data validity | Cognitive synthesis and critique |
| Purpose | Aggregation of evidence | Structuring and expansion of knowledge |

In the future, this comparison and juxtaposition will help understand the value and necessity of the PANDAVA protocol in scientific research and beyond.

Accordingly, PANDAVA is particularly needed and can be used in theoretical fields (philosophy of science, information theory, cognitive science, systems evolution), interdisciplinary research (biosynthesis, biophysics, epistemology of digital systems), and emerging disciplines. This is precisely where there are no established empirical datasets yet and where semantic gaps are enormous.

Moreover, PANDAVA can be used when there is a need to create new hypotheses (Nickerson, 1998), particularly in projects where not only a review but also the development of a paradigm is expected.

**2. Methodology**

This study is based on the development and testing of a new systematic knowledge analysis protocol, PANDAVA (Protocol for Analysis and Navigation of Deep Argumentative and Valued Knowledge).

The methodology combines the principles of traditional systematic review protocols, such as PRISMA, with original cognitive and semantic analysis procedures aimed at mapping semantic structures, identifying epistemic gaps, and generating new hypotheses.

The study employed semantic annotation of theoretical concepts, clustering of concepts on the basis of maturity and connectedness criteria, construction of heatmaps of semantic gaps, numerical modelling of the epistemic landscape, and scenario-based synthesis of knowledge on the basis of the identified connections and contradictions.

The methodology is oriented towards theoretical–analytical application in interdisciplinary scientific fields, where traditional systematic review models prove insufficient.

As part of the implementation of the PANDAVA protocol, the following digital tools and libraries were used:

| Stage | Tools |
| --- | --- |
| Working environment | Jupyter Notebook |
| Semantic extraction of concepts | Python, libraries: SciSpacy |
| Clustering | Python (KMeans, PCA) |
| Gap Mapping | Python (Matplotlib), Seaborn |
| 3D Visualization | Matplotlib |
| Concept network construction | NetworkX |
| Scenario synthesis | Combined heuristics |

The following key variables were used in the study:

| Variable | Description | Range |
| --- | --- | --- |
| Ontological Clarity (OC) | Degree of ontological clarity of the concept | 0–3 points |
| Argumentative Depth (AD) | Saturation of the concept with argumentation | 0–3 points |
| Theoretical Coherence (TC) | Integration of the concept into theoretical frameworks | 0–3 points |
| Generativity (G) | Ability of the concept to generate new ideas | 0–3 points |
| Epistemic Robustness (ER) | Concept's resistance to criticism | 0–3 points |

The final overall maturity of a concept was calculated as the sum of the five variables (maximum 15 points).

Numerical modelling was conducted to analyse the epistemic landscape and identify clusters of mature concepts, areas of semantic gaps, and logical trajectories of hypothesis generation.

The main modelling stages included
1. *PANDAVA-Eval Matrix:*
   Evaluation of each concept across five maturity criteria.
2. *Gap Matrix:*
   Inversion of maturity assessments to construct a map of semantic gaps.
3. *PCA:*

Dimensionality reduction of the data is used to construct a two-dimensional map of concepts.
4. K-Means clustering:
The concepts are grouped into three main semantic clusters: mature concepts (core), intermediate concepts (bridges) and problematic concepts (risk zones).
5. *3D Visualization:*
Construction of a three-dimensional epistemic landscape of gaps via surface plotting in Matplotlib.

The procedure for conducting the study was interpreted through the stages of applying the protocol

| № | Stage | Module Name | Purpose | Tools/Methods |
|---|---|---|---|---|
| 1 | P1 | Problem Cartography | Construction of an ontological map of the studied problem | Mind maps (XMind) |
| 2 | A1 | Argument Harvesting | Extraction of arguments, models, hypotheses from the literature corpus | SciSpacy |
| 3 | N1 | Navigation through Knowledge Structures | Construction of a network of semantic and logical connections between concepts | NetworkX |
| 4 | D1 | Deep Clustering | Clustering of concepts based on maturity and structure (Grant, 2009) | KMeans, PCA, Radar Chart, Heatmap |
| 5 | A2 | Argumentative Gap Mapping | Identification of semantic gaps (Firestone, 1993) in the theoretical structure | 3D Epistemic Landscape (Matplotlib) |
| 6 | V1 | Value-Oriented Synthesis | Synthesis of concepts into a holistic knowledge architecture | XMind, synthesis graphs, flowcharts |
| 7 | A3 | Applied Hypothesis Generator | Generation of new scientific hypotheses based on identified gaps and connections | Heuristic templates |

For verification of the PANDAVA protocol, the following criteria were applied:

| Criterion | Description |
|---|---|
| Reproducibility | The ability to repeat the stages of concept evaluation and map construction |
| Semantic completeness | Completeness of coverage of key concepts, models, arguments |
| Epistemic diagnosticity | Protocol's ability to identify weak areas and potential directions for synthesis |
| Generative productivity | Number of new hypotheses generated through the protocol |

The following methodological limitations should be noted:
- the protocol requires significant interpretive work during the concept annotation stage,
- some elements of evaluation (e.g., maturity) may include a degree of subjectivity, minimized through double independent expert evaluation,
- The application of these methods to empirical studies requires the adaptation of maturity criteria to the specifics of experimental data.

The research variables are interpreted in the following technical table:

| № | Variable | Description | Type | Value Range |
|---|---|---|---|---|
| 1 | Ontological Clarity (OC) | Clarity of defining entities and relations in the concept | Evaluation scale | 0–3 |
| 2 | Argumentative Depth (AD) | Degree of saturation of the concept with logical argumentation and examples | Evaluation scale | 0–3 |
| 3 | Theoretical Coherence (TC) | Integration of the concept into existing theoretical frameworks | Evaluation scale | 0–3 |
| 4 | Generativity (G) | Ability of the concept to generate new ideas and directions | Evaluation scale | 0–3 |
| 5 | Epistemic Robustness (ER) | Resistance of the concept to criticism, alternative explanations, and new data | Evaluation scale | 0–3 |
| 6 | Summed Concept Maturity (SCM) | Sum of assessments across all five maturity criteria | Calculated variable | 0–15 |
| 7 | Gap Score (GS) | Inverted maturity score for displaying gaps | Calculated variable | 0–3 (for each criterion) |

The variables were applied to each unit of analysis (concept) to evaluate its maturity, connectedness, and resilience.

The stages were sequentially applied within the PANDAVA protocol, passing through all phases from problem cartography to the generation of new hypotheses.

## 3. Results

### *3.1. Structure of PANDAVA*

The PANDAVA protocol (Protocol for Analysis and Navigation of Deep Argumentative and Valued Knowledge) was developed as an alternative to linear, empirically oriented methodologies of systematic reviews. Its main goal is not only to aggregate results but also to uncover the logical-semantic structures of scientific texts, identify epistemic clusters, and form trajectories for intellectual synthesis (Table 5).

PANDAVA is aimed at a systematic, philosophically grounded, and instrumentally realizable investigation of the conceptual depth and argumentative architecture of a scientific field.

**Table 5.** Modules of the PANDAVA Protocol (7 stages)

| № | Module | Purpose |
|---|---|---|
| 1 | P1. Problem Cartography | Ontological and conceptual mapping of the studied problem |
| 2 | A1. Argument Harvesting | Semantic extraction of concepts, theories, models, types of arguments |
| 3 | N1. Navigation through Knowledge Structures | Identification of levels of analysis, ontological layers, and conceptual connections |
| 4 | D1. Deep Clustering | Epistemic clustering: by type of knowledge, maturity, connectedness |
| 5 | A2. Argumentative Gap Mapping | Identification of gaps, contradictions, epistemic ruptures |
| 6 | V1. Value-Oriented Synthesis | Integration of theories, synthesis of directions, constructive reinterpretation |
| 7 | A3. Applied Hypothesis Generator | Formulation of new hypotheses, concepts, and research scenarios |

Each module corresponds to a set of techniques, criteria, and visual outputs, such as conceptual graphs, gap-radar diagrams, and semantic maps (Table 6).

**Table 6.** Instrumental Stack

| Stage | Approach/Technology |
|---|---|
| Concept extraction | NLP |
| Clustering | Semantic embeddings + manual classification |
| Gap analysis | Radar diagrams (Plotly, Matplotlib) |
| Visualization | NetworkX |
| Scenario synthesis | Concept combinations + heuristic rules |
| Hypothesis generation | Templates: if A ←→ B + gap → propose C |

Thus, the protocol is implemented as a flexible cognitive route in which the researcher can:
- *start by framing the problem in an ontological context (P1),*
- *Load a corpus of scientific texts (articles, preprints, case studies),*
- *annotate the main theoretical-semantic units via NLP tools or manually (A1),*
- *Build a network of conceptual connections and visualize it as a cognitive map (N1),*

*- perform clustering on the basis of epistemic features (D1), including elements such as theoretical nature, maturity level, explanatory power, and contradictions;*
*- form a gap radar (A2) and identify zones where knowledge is lacking, hypotheses contradict each other, or connectedness between theoretical levels is missing;*
*- move to scenario synthesis (V1), proposing bridges between concepts and predicting directions of idea convergence;*
*- Finally, formulate higher-level hypotheses (A3) on the basis of the intersection of weakly connected concepts.*

At each stage, the protocol provides a reflection of the researcher's position.

In this context, the selection of concepts is carried out with acknowledgement of interpretative subjectivity, interpretations are recorded and commented upon, and multiple epistemic maps based on different strategies of understanding are allowed.

### 3.2. Modules of PANDAVA

Figure 1 presents a visual diagram of the PANDAVA protocol modules, which is structured as a sequential, cognitive-semantic chain from problem cartography (P1), through the collection and structuring of arguments (A1 → N1), to clustering and gap identification (D1 → A2), and further to knowledge synthesis and the generation of new hypotheses (V1 → A3).

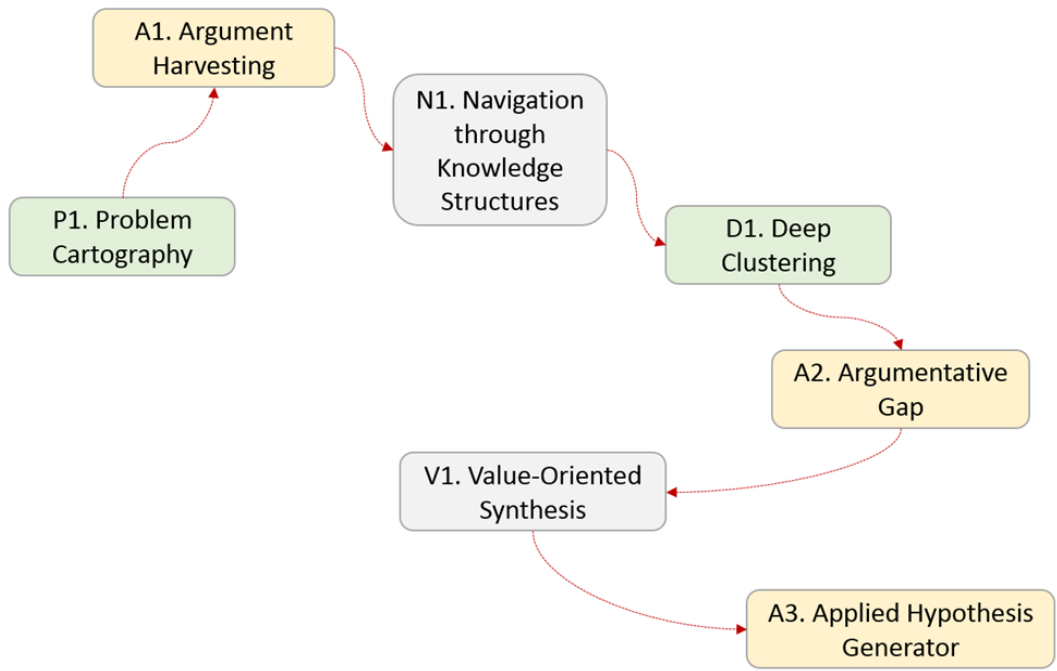

**Figure 1.** Diagram of the PANDAVA Protocol Modules

Here, module P1. Problem cartography reflects the ontological mapping and framing of a problem. Module A1. Argument harvesting and N1. Navigation through knowledge structures interprets semantic and structural analyses. Modules D1. Deep Clustering and A2. Argumentative gap mapping is responsible for identifying epistemic gaps and conflicts. Module V1. Value-Oriented Synthesis and A3. The applied hypothesis generator corresponds to knowledge synthesis and hypothesis generation.

The presented module diagram of the protocol is not built as a linear bureaucratic procedure but rather as an evolutionary cognitive transformation of the scientific field. Each step is a semantic operation, transforming not data but conceptual structures.

*P1. Problem Cartography*

This is not simply posing a question but creating a semantic map of the studied area. The researcher must identify key concepts, levels of analysis (individual, model, system), and contexts (scientific schools, epistemological frameworks).

Thus, P1. Problem cartography is interpreted as the phase of ontological framing and semantic localization.

*A1. Argument Harvesting*

This is the stage of collecting not facts but arguments, hypotheses, and models. Methods of NLP annotation, manual decomposition of publications, and logical-semantic extraction are used.

In fact, it is analogous to "*semantic mining*," where ideas are extracted from texts.

*N1. Navigation through Knowledge Structures*

Here, we construct a network of semantic connections between arguments. The following levels are identified: explanation vs. prediction, theory vs. model, counterarguments, and paradoxes.

In this sense, N1. Navigation through knowledge structures is the "navigation" part, which involves understanding the topology of knowledge.

*D1. Deep Clustering*

At this stage, concepts are grouped by type of knowledge (theory, method, critique), maturity, connectedness, and conflict.

As a result, an epistemic map is formed—not by topics but by the depth and nature of knowledge.

*A2. Argumentative Gap Mapping*

This is the most important stage of analysis: where is justification missing? where do approaches conflict? Which zones remain theoretically "invisible"?

At this stage, Gap-Radar is created: a map of gaps, overloaded zones, and unmarked clusters.

*V1. Value-Oriented Synthesis*

Here, cognitive synthesis is interpreted, meaning the connection of incompatible models, the identification of latent bridges, and the rethinking of conflicts as zones of growth.

This stage of constructive thinking is analogous to intellectual design thinking.

*A3. Applied Hypothesis Generator*

This is the final stage, where higher-level hypotheses are formulated. They are not generated from general reading but from the intellectual analysis of the structure of knowledge itself.

The visual structure of A3. The applied hypothesis generator represents the map of the transition from knowledge to further knowledge.

Thus, in the context of the PANDAVA protocol module diagram, knowledge is not a list but a structure.

In this context, PANDAVA does not simply classify sources but rather structures meaning. In PANDAVA, a review becomes a process of thinking. Instead of describing the literature, we engage in redefining the scientific situation.

Each module is interpreted as a separate cognitive operation. This is not mere formality but a phase of thinking—the transition from cartography to argumentation, navigation, and synthesis.

Thus, the PANDAVA scheme represents an intellectual trajectory along which the researcher moves, a graph of cognitive operations leading to the creation of new knowledge and an alternative to formalism.

It is a path where a hypothesis is born from comprehension, not merely from summation.

### 3.3. PANDAVA-Eval Matrix

Let us consider the PANDAVA-Eval matrix for assessing the maturity, depth, and value of scientific concepts extracted during the semantic review process via the PANDAVA protocol.

The matrix represents a universal scale of five cognitive axes (Table 7) that help the researcher formally evaluate the quality and density of a concept, compare concepts within one field or across disciplines, select which concepts to include in synthesis or clustering, and substantiate conclusions and hypotheses on the basis of epistemic evaluations.

**Table 7.** PANDAVA-Eval Matrix: Concept Evaluation Scale

| № | Criterion | Description | Evaluation (0–3) |
|---|---|---|---|
| 1 | Ontological Clarity | How clearly is the concept defined? Does it have clear boundaries and internal structure? | 0 = vague, 3 = strictly defined |
| 2 | Argumentative Depth | Is there a logical structure, a system of arguments, examples, a model? | 0 = declarative, 3 = evidential |
| 3 | Theoretical Coherence | How well is the concept embedded in existing theories, paradigms, or explanatory frameworks? | 0 = isolated, 3 = highly integrated |
| 4 | Generativity | How much does the concept generate new ideas, hypotheses, research directions? | 0 = inert, 3 = productive |
| 5 | Epistemic Robustness | Is it supported in the literature, reproducible, and does it possess scientific "survivability"? | 0 = single mention, 3 = widely cited, adaptable |

The maturity level is interpreted through a corresponding scale (Table 8).

**Table 8.** Scale: from 0 to 15 points

| Points | Maturity Level | Interpretation |
|---|---|---|
| 0–4 | Immature Concept | Requires rethinking or exclusion |
| 5–9 | Transitional/Limited | Can be used with reservations |
| 10–12 | Constructive | Suitable for synthesis and discussion |
| 13–15 | Core Concept | Candidate for key nodes of knowledge (core theory) |

The matrix in PANDAVA can be applied at stages D1 (deep clustering) and A2 (gap mapping), helping to visualize the distribution of maturity through histograms, radar charts, and heatmaps. It is suitable for justifying the inclusion or exclusion of concepts and can be included in the report as an expert panel for evaluating conceptual value.

To illustrate, let us present an approximate visual template of the PANDAVA-Eval matrix as follows:

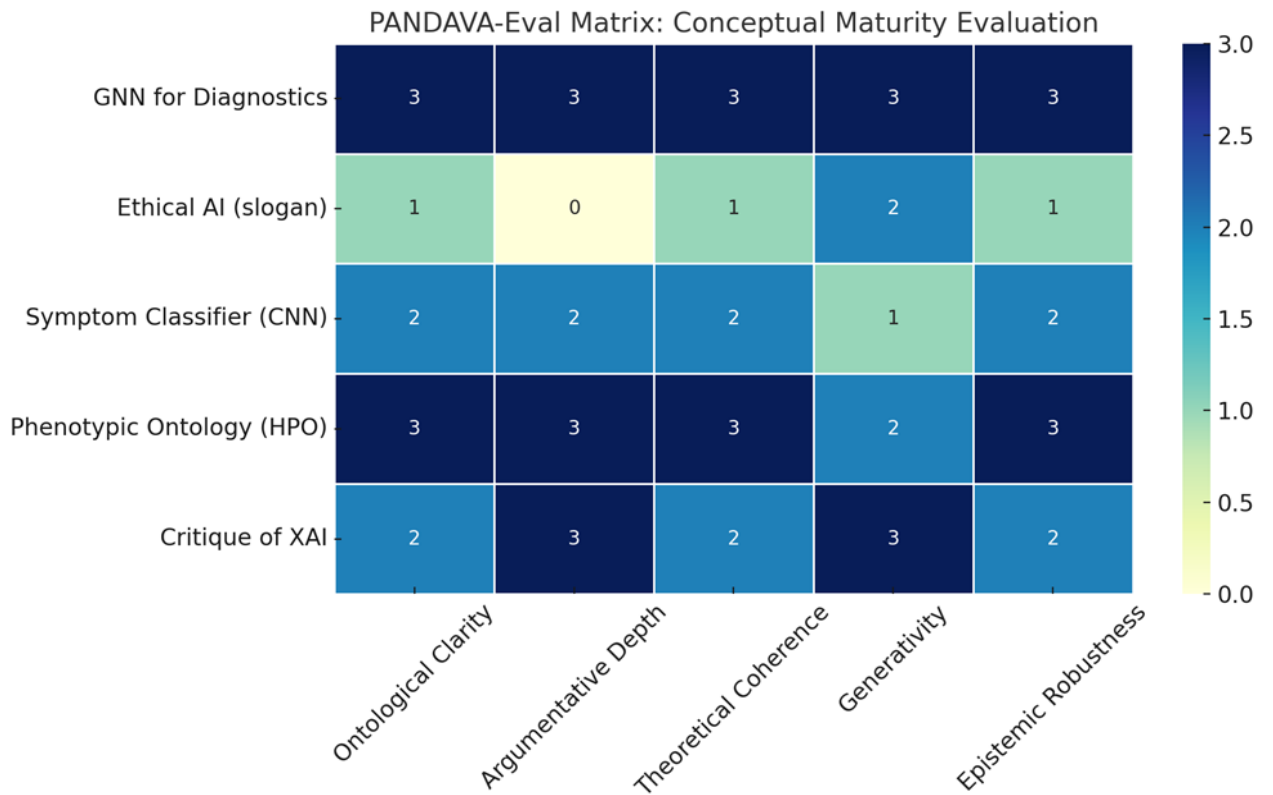

**Figure 2**. Visual template of the PANDAVA-Eval matrix

The visual template of the PANDAVA-Eval matrix represents a heatmap assessing the maturity and significance of concepts across five key criteria (Figure 2):
- *Ontological clarity as conceptual precision,*
- *Argumentative depth as the strength of argumentation,*
- *Theoretical coherence as a connection with theories,*
- *Generativity as the productivity for generating new ideas,*
- *Epistemic robustness as resilience in scientific discourse.*

Essentially, the PANDAVA-Eval matrix interprets the epistemic value of the concepts extracted during the review study following the PANDAVA protocol.

The visualization clearly displays the distribution of the cognitive and ontological qualities of the concepts across five key criteria. Each criterion assesses a specific aspect of scientific maturity, starting from conceptual clarity and ending with the concept's productivity in generating new knowledge.

In the context of the matrix, even within a single subject area (e.g., AI or medicine), concepts can vary significantly in depth, resilience, and epistemic usefulness.

This not only allows for systematization of the field but also enables the prioritization of synthesis: what to include, what to question, and what to develop.
Overall, the application of the PANDAVA-Eval matrix shows that a systematic review becomes not only a collection of facts but also an evaluation of the conceptual quality of science and a means of theoretical sense-making, transcending mere frequency analysis.

Thus, the matrix confirms the value of PANDAVA as a reflective and meta-analytical protocol that enables the quantitative and visual recording of knowledge structures, the identification of cognitive distortions (e.g., dominance of slogans over theories), and the making of reasoned decisions during the synthesis process.

### *3.4. Methodology for Evaluating Concept Maturity*

The methodology is intended for the systematic and reproducible evaluation of the maturity of theoretical concepts within the framework of the PANDAVA protocol.

It enables the identification of strengths and weaknesses in a research field, the structuring of conceptual maps, the conscious synthesis of knowledge, and the minimization of the inclusion of weak or pseudotheoretical ideas. Each concept is evaluated on the basis of five criteria reflecting its epistemic value (Table 9).

**Table 9.** Five Criteria for Concept Evaluation

| Criterion | Evaluation Content |
|---|---|
| Ontological Clarity | Clarity and definiteness of entities and their relations within the concept |
| Argumentative Depth | Depth, rigor, and completeness of the concept's argumentation |
| Theoretical Coherence | Integration of the concept into existing scientific theories |
| Generativity | The ability of the concept to generate new ideas and expand the research field |
| Epistemic Robustness | Resistance of the concept to criticism, alternatives, and new data |

Each criterion is assessed on a four-level scale

**Table 10.** Four-level scale

| Score | Interpretation |
|---|---|
| 0 | No features present |
| 1 | Features are weakly expressed, vague |
| 2 | Features are present but incomplete |
| 3 | Features are clearly, fully, and stably expressed |

Notably, all the criteria must be evaluated independently of one another. It is acceptable for a single concept to demonstrate different levels of maturity across different criteria.

The evaluation procedure involves the following sequential stages:

1. *Ontology Analysis of the Concept:*
   Identification of how clearly basic elements (entities, processes, relations) are defined.
2. *Argumentation Analysis:*
   Checking for the presence of rigorous logical structures, experimental confirmations, or well-substantiated models.
3. *Verification of Theoretical Coherence:*
   Assessment of how well the concept is integrated into existing scientific paradigms or properly justifies the need for their revision.
4. *Evaluation of generative potential:*
   Examination of whether the concept can generate new studies, models, or hypotheses.
5. *Verification of Epistemic Robustness:*
   Analysis of how well the concept withstands criticism, alternative explanations, and new data.

The principles of ensuring objectivity in the evaluation can be formulated through three procedures:

- The evaluation must be conducted by two or more independent experts, with subsequent reconciliation of discrepancies.
- If necessary, extended checklists can be used for each criterion (e.g., "*Is there a strict definition? *", "*Has the concept been tested against alternative data?*" etc.).
- 

Thus, the a priori evaluation of concept maturity in PANDAVA enhances the cognitive rigor of systematic reviews, reveals the real structure and weak zones of the theoretical field, creates a foundation for the qualitative synthesis of knowledge, and promotes the formulation of new scientific hypotheses on the basis of a solid foundation.

Accordingly, the PANDAVA-Eval matrix turns literature analysis into a full-fledged epistemic audit of the research space.

The standardized template for the PANDAVA-Eval matrix form can be formulated through the following sequence of actions:

1. Select a concept (theoretical model, hypothesis, idea),
2. Each of the five maturity criteria is evaluated on a scale from 0 to 3: 0 = feature absent, 1 = feature weakly expressed, 2 = feature moderately expressed and 3 = feature strongly expressed.
3. Justify your evaluation with a brief explanation (1–2 sentences),

4. Result: The concept receives its "maturity profile" for further clustering or analysis.

The structure of the form for a single concept (Table 11) is as follows:

**Table 11.** Structure of the Form for a Single Concept

| № | Criterion | Evaluation (0–3) | Justification |
|---|---|---|---|
| 1 | Ontological Clarity | | How clearly are entities and relations defined in this concept? |
| 2 | Argumentative Depth | | How deep and rigorous is the argumentation? |
| 3 | Theoretical Coherence | | How well is the concept integrated into the existing theoretical system? |
| 4 | Generativity | | Can the concept generate new ideas or research directions? |
| 5 | Epistemic Robustness | | How resistant is the concept to criticism and alternative explanations? |

The features of the template ensure that

- *transparency (each evaluation requires a justification),*
- *reproducibility (with multiple assessments by different researchers, scores can be compared),*
- *flexibility (can be used both in manual evaluation and in semiautomated systems).*

### *3.5. Gaps for Concepts*

The Gap Heatmap for concepts following the PANDAVA protocol (Figure 3) shows the intensity of the color scale; the more intense the color is, the stronger the semantic or epistemic gap for a specific criterion:

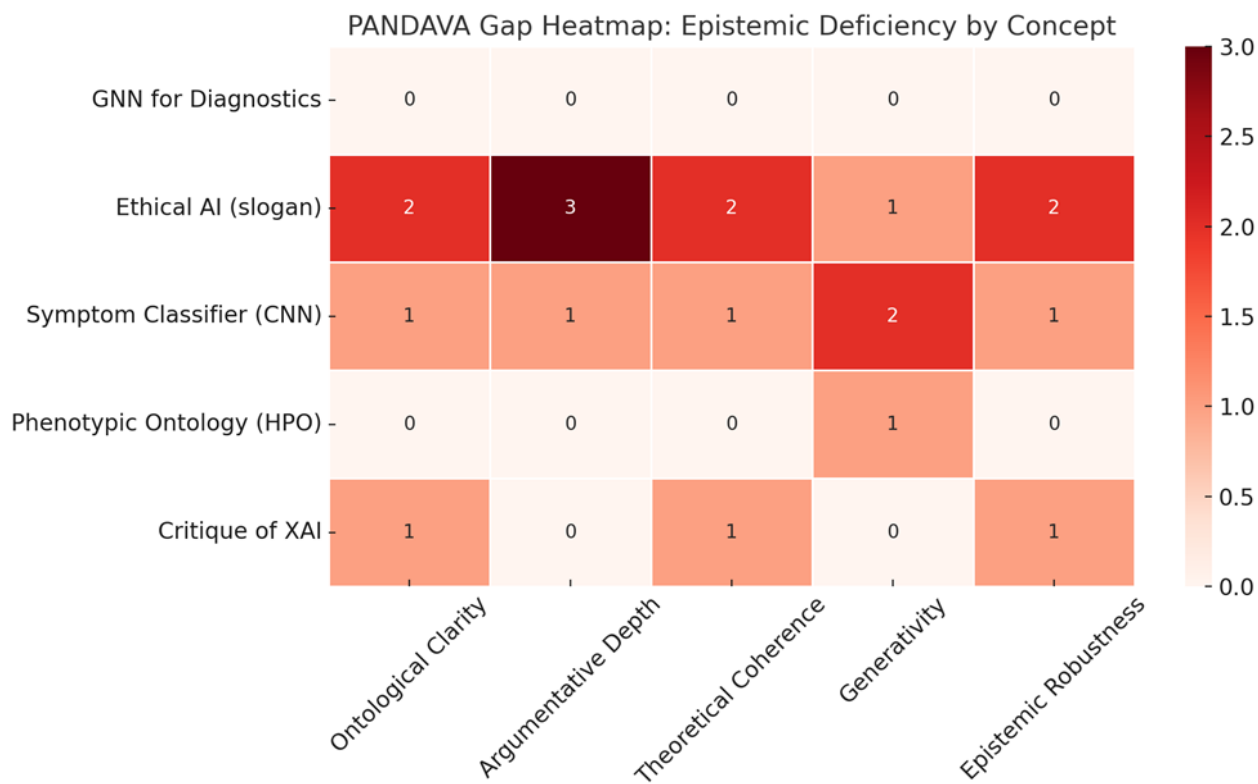

**Figure 3.** Gap Heatmap for Concepts According to the PANDAVA Protocol

Here, ethical AI (slogan) demonstrates the largest gaps across almost all axes. The symptom classifier (CNN) shows gaps in generativity and theoretical coherence. GNN and HPO display minimal gaps, especially in terms of argumentation and robustness.

The Gap Heatmap is not merely a technical visualization; it reflects deep semantic and structural ruptures in scientific knowledge.

In the PANDAVA protocol, a gap is understood as an epistemic weakness or deficiency of a concept in relation to a specific maturity criterion (Table 12). It is not just the absence of information but a structural vulnerability of the concept, manifesting as follows:

**Table 12**. Gap criteria

| Criterion | A gap means... |
|---|---|
| Ontological Clarity | The concept is vague, metaphorical, lacks clear boundaries. Often used as a slogan or "buzzword." |
| Argumentation | The concept is not accompanied by evidence, logical structure, or examples. It remains a bare assertion. |
| Coherence | The concept is not integrated into theories, models, or scientific paradigms. It "stands aside." |
| Generativity | The concept does not produce new hypotheses, ideas, or research directions. It is "dead-end." |

| Epistemic Robustness | The concept is not anchored in science: it is not used, reproduced, or adapted. |

According to the table, examples of gap interpretations are as follows:

*Ethical AI (slogan):*
Gap in argumentation: This gap lacks logic, which is often used declaratively.
Gap in coherence: not integrated into formal models or computational practices.
Interpretation: the concept requires reformulation or operationalization. Otherwise, it will hinder rather than assist scientific synthesis.
*Symptom Classifier (CNN):*
Gap in generativity: technically important but does not generate new theories.
Interpretation: applicable as an auxiliary tool but not as a core concept for synthesis.
*GNN for diagnostics and HPO:*
Minimal gaps: robust, formal, theoretically connected, and productive.
Interpretation: These are key structural concepts that can serve as anchors for building new hypotheses and synthetic models.

Notably, traditional protocols do not capture such gaps because they evaluate articles rather than concepts, focus on quantitative metrics rather than semantic structures, and ignore the conceptual contradictions and weaknesses of theoretical foundations.

Moreover, the PANDAVA protocol, for the first time, enables the detection of semantic "holes" before building synthesis, distinguishes between mature and immature knowledge, and visualizes the structure of epistemic distortions.

Depending on the nature of the gaps, appropriate actions are taken (Table 13).

**Table 13.** What to do with Gaps?

| Gap Level | Recommendation |
|---|---|
| Critical | Exclude the concept or subject it to revision |
| Moderate | Use with refinement or clarification |
| Minor | Include in the synthesis |

The gaps are interpreted in the form of an epistemic landscape (3D diagrams and cluster maps) (Figure 4).

The epistemic landscape is a 3D model of the distribution of semantic gaps, and its visual representation serves as a tool for epistemic diagnostics of the knowledge field—identifying its weak, conflictual, and potentially evolving zones.

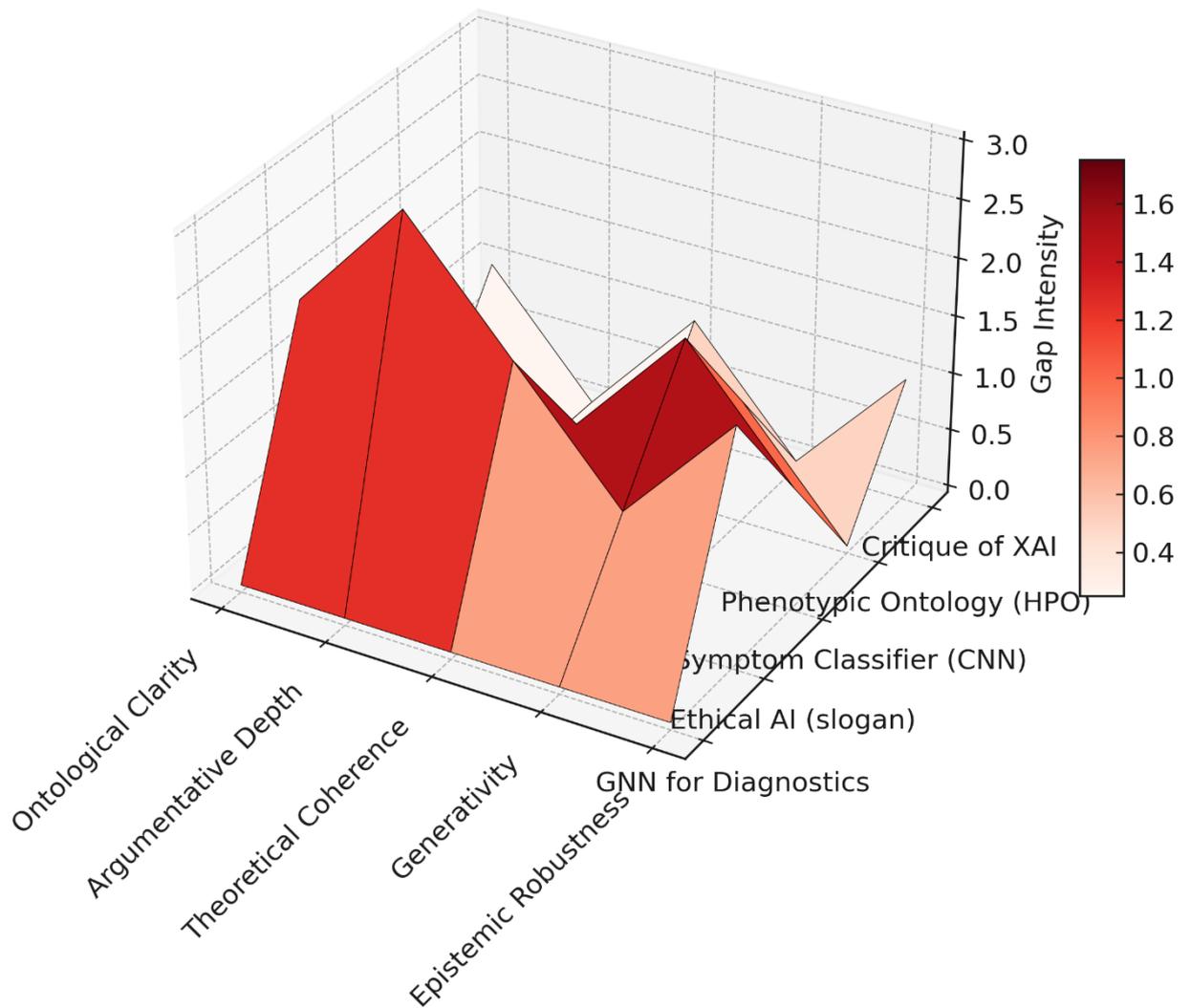

**Figure 4.** 3D visualization of the epistemic gap landscape (PANDAVA Gap landscape)

On this chart, the X-axis represents the key maturity criteria (ontology, argumentation, etc.), the Y-axis lists the analysed concepts, and the Z-axis shows the gap intensity (0 = mature, 3 = maximum deficiency; the higher the value is, the worse).

The peaks on the graph indicate the most critical ruptures in understanding the scientific robustness of the concepts. Flat regions represent balanced and mature concepts.

*Peak Gap Zones (according to the graph)*
Ethical AI (slogan):
The highest peaks in argumentation (3), ontology (2), and robustness (2). A typical example of a highly cited yet theoretically empty concept. Its usage leads to fragmentation of the field, as it provides no theoretical foundation.
Recommendation: requires strict operationalization or exclusion from the synthesis core.

Symptom Classifier (CNN):

Local peaks in generativity (2) and coherence (1). Low conceptual productivity—more of a tool than an idea.

Recommendation: suitable as a supporting element but not as a conceptual center.

*Maturity plateaus (according to the graph)*

GNN for diagnostics:

Nearly perfect flat line (all gaps = 0). Robust, well-argued, reproducible, and theoretically embedded.

Recommendation: use as a cognitive anchor point when constructing hypotheses.

Phenotypic Ontology (HPO):

Small local rise in generativity. Otherwise, it is a mature and well-structured concept.

Recommendation: use as a foundational ontological base for clustering.

*Complex Contour* - Critique of XAI:

Wavelike landscape: This landscape is strong in argumentation but lacks coherence and robustness. This is a controversial but necessary concept — a zone of cognitive friction.

Recommendation: do not exclude but include as a reflective axis for synthesis.

Importantly, the presence of a gap does not imply that the concept is useless. In contrast, it may represent a growth point — if developed, refined, or embedded in a model. Instead of a binary "include/exclude" decision, we obtain a topology of the semantic field, where bridges can be built between concepts, risks and potentials of integration can be assessed, and we can visually detect where knowledge is incomplete.

Thus, PANDAVA functions as an architectural protocol, where the landscape is evidence that PANDAVA is not merely a review but also a tool for the intellectual design of science.

The protocol is an epistemic architecture, not a data aggregator.

The 3D epistemic landscape therefore provides a deep meta-map of the knowledge field, allowing the identification of critical zones, the formation of architectures for future research, and support for conceptual coherence.

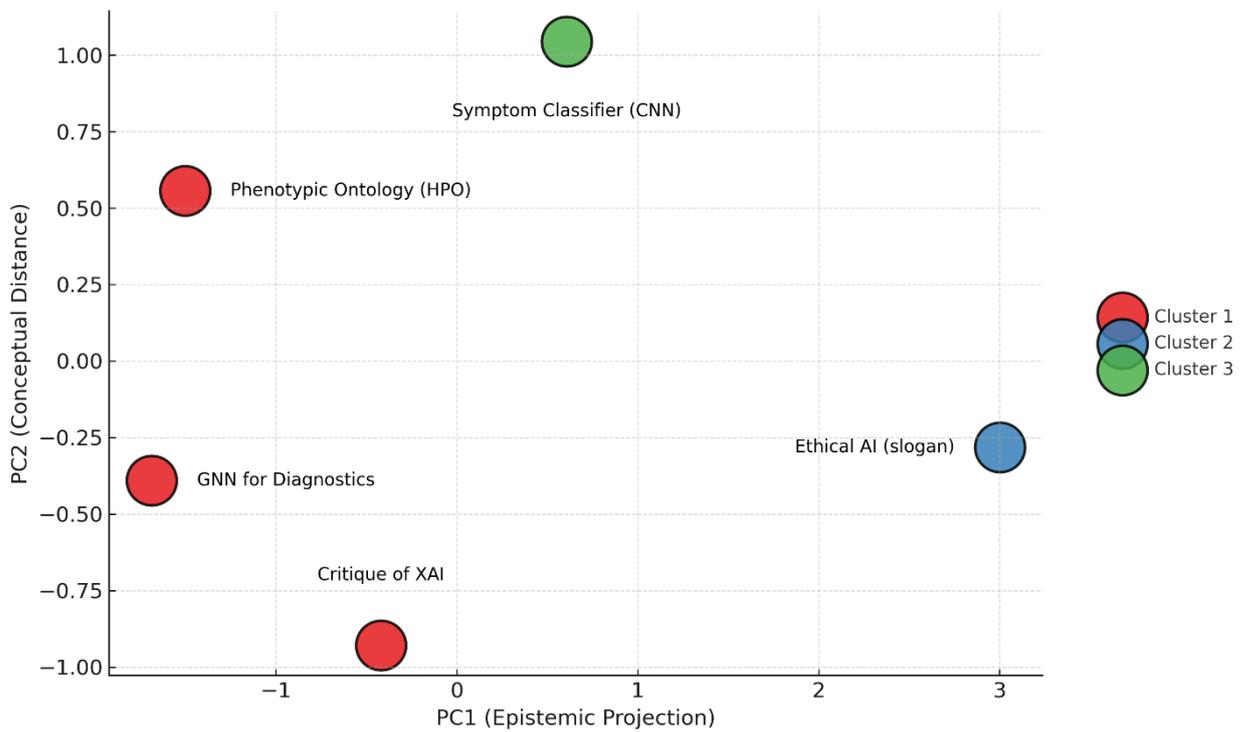

**Figure 5.** Epistemic Gap Clustering Map (PANDAVA Epistemic Gap Clustering)

Here, concepts are grouped into three clusters on the basis of the similarity of their gaps. Placement along the axes reflects semantic and argumentative proximity. One cluster includes mature concepts (e.g., GNN and HPO). Another brings together weak and declarative ideas (e.g., ethical AI). The third cluster consists of intermediate, applied, yet not fully theorized, concepts.

Figure 5 presents a two-dimensional projection of the epistemic gaps between concepts (via PCA) followed by k-means clustering (3 clusters, see Table 14). The PC1 axis (epidemic projection) shows the main direction of variation in the gaps between concepts. The PC2 axis (Conceptual Distance) reflects secondary differences in the depth and breadth of semantic deficiencies.

**Table 14.** Cluster Interpretation

| Cluster | Composition | Characteristics |
| --- | --- | --- |
| Cluster 1 | GNN for Diagnostics, Phenotypic Ontology (HPO) | Minimal gaps, high maturity |
| Cluster 2 | Critique of XAI, Symptom Classifier (CNN) | Moderate gaps, partial constructiveness |
| Cluster 3 | Ethical AI (slogan) | Maximum gaps, declarative nature |

Clusters GNN for diagnostics and Phenotypic Ontology (HPO) are compactly located in the right part of the graph, with minimal gaps across all key criteria (ontology, argumentation, coherence, generativity, robustness). These concepts can be used as the core framework of synthetic knowledge within the domain.

Characteristics: high ontological clarity and argumentation, well integrated into theoretical and methodological landscapes, serve as anchor concepts for building synthesis, forecasting, and reconfiguration of knowledge. Roles in PANDAVA: foundational for V1 (value-oriented synthesis) and A3 (hypothesis generator); core for forming cognitive clusters of new knowledge.

The cluster symptom classifier (CNN) and XAI criterion are located closer to the center but in different quadrants. These display moderate gaps: CNNs are limited in generativity, and XAI critique is rich in argumentation but not fully integrated into the ontology of applied models. Thus, these concepts are suitable as bridges or supporting structures for enriching synthesis.

Characteristics: These characteristics have practical value but are limited in ontological and generative roles.

The XAI technique is argumentatively rich but unstable; CNNs are technically important but do not generate ideas. Roles in PANDAVA: suitable for A1 (Harvesting) and D1 (Clustering), but not for direct hypothesis generation; can be integrated through V1 scenarios as secondary elements.

The cluster Ethical AI (slogan) is located on the left side of the graph and is separated from the others. It demonstrates maximum epistemic vulnerability: weak argumentation, vague ontology, and insufficient theoretical coherence.

Characteristics: slogan-like, vague, poorly argued; widely used in discourse but lacking theoretical substance and prone to semantic inflation (used without comprehension).

At the same time, this cluster is in a risk zone: substituting content with form, potentially misleading the synthesis and generating pseudocoherence. Role in PANDAVA: monitored in A2 (gap mapping) as a problematic zone requiring either clarification or exclusion.

Summary of Cluster Meanings
- mature concepts are knowledge anchors, stabilizers, and foundations for synthesis.
- Gap-laden concepts are indicators of voids and the pseudointellectual context. Intermediate concepts act as bridges between pragmatics and theory and are useful for synthesis refinement.

Key Observations from the Graph

1. *Navigation through the semantic field:*
   The graph allows quick identification of concepts as cores of future synthesis, support bridges, potential sources of error, or pseudoscientific inclusions.
2. *Architecture of scientific construction:*
   Clusters help plan which ideas to include without changes, which to revise, and which to exclude or critique.

3. *Developmental trajectories:*
    It is possible to design paths for concept improvement via strengthened argumentation, increased ontological precision, and elimination of semantic gaps.

In our view, the PANDAVA epidemic gap clustering graph is an effective tool for intellectual navigation throughout the research field. It allows us to see not only facts but also the logical fabric of the scientific landscape, revealing risk zones and synthesis opportunities. The application of such maps significantly improves the quality of hypotheses, depth of reviews, and structural coherence of scientific research.

### *3.6. Multilevel Synthesis Map*

Figure 6 presents a multilevel knowledge synthesis map according to the PANDAVA protocol in the form of a semantic vertical.

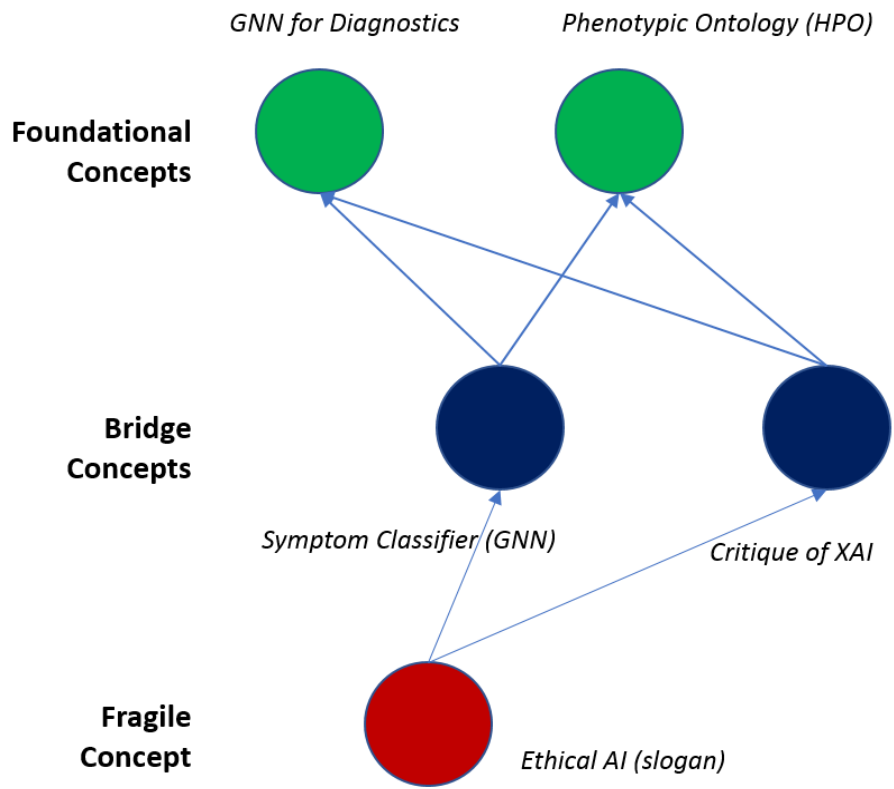

**Figure 6.** Multilevel Synthesis Map

Foundational concepts and phenotypic ontology (HPO) serve as the basis for theoretical construction and hypothesis formation. Bridge concepts are interpreted as connectors and intermediaries; *the symptom classifier (GNN)* and *the XAI* criterion link the applied and theoretical levels. The fragile concept represents a semantic *risk—* *ethical AI (slogan)* acts more as a signal of cognitive instability.

The arrows in the diagram show how low-structured concepts can be transformed into more mature levels through cognitive synthesis.

Thus, the multilevel synthesis map represents an architecture of knowledge, not just a collection of scientific content.

On the basis of the above and the synthesis map, we can more consistently interpret the step-by-step application of the protocol in the context of the overall picture (Table 15).

**Table 15.** Step-by-step instructions for applying the PANDAVA protocol

| Target | Actions | Tools | Result |
|---|---|---|---|
| colspan Step 1 — P1: Problem Cartography ||||
| to form an ontological field of research | Define the subject, levels of analysis (individual, model, system), type of knowledge (theory, model, method). Build a semantic map of the concepts included in the problem. Divide key concepts into: explanatory, operational, conflicting | Mind maps (XMind, Miro, CmapTools)<br><br>Category Table (Excel/Sheets) | Ontological map of the region<br><br>List of key concepts to extract |
| colspan Step 2 — A1: Argument Harvesting ||||
| annotate and structure concepts from the literature | Collect a relevant literature corpus (according to the inclusion criteria). For each source, select: central hypothesis, model (if any), the main argument, conceptual novelty | Annotation table (automated or manual)<br><br>NLP | Catalogue of concepts<br><br>Primary Maturity Table |
| colspan Step 3 — N1: Navigation through Knowledge Structures ||||

| build a network of logical and semantic connections between concepts | Link concepts by type: Justification → Theory Controversy → Conflict Support → Consent Build a graph of interactions between concepts | NetworkX/Gephi/yEd  Neo4j (if storage is needed) | Conceptual graph  Visualization of logical-semantic structure |
|---|---|---|---|

Step 4
D1: Deep Clustering

| classify concepts by maturity, coherence, productivity | Apply PANDAVA- Eval Matrix to each concept: ontology, argumentation, connectivity, generativity , stability. Calculate the total score. Divide concepts into: nuclear, constructive, limited, problematic | Excel Table/Sheets  Python/Colab template (automatic visualization)  Radar Chart/Heatmap | Matrix of maturity  Clusters of semantic value |
|---|---|---|---|

Step 5
A2: Argumentative Gap Mapping

| identify semantic gaps and unfilled areas of knowledge | Invert the maturity matrix (3 - value) → get a gap map. Build a heatmap and 3D epistemic landscape. Find: zones of logical conflict, missing ligaments, paradoxical intersections | Heatmap (seaborn), 3D plot (matplotlib)  PCA + KMeans → gap cluster map | Gap Radar  Epistemic Landscape  Clusters of Contradictory Concepts |
|---|---|---|---|

Step 6
V1: Value-Oriented Synthesis

| create a cognitive model or theoretical framework that connects key concepts | Combine core and intermediate concepts into a single map. Build a multilayered knowledge architecture: basis, | Scheme « Synthesis Map »  Ontographs , Sankey diagrams, flow charts | The holistic structure of synthesized knowledge |
|---|---|---|---|

| | bridges, restrictions. Apply reflection: how will synthesis affect the paradigm? | | Preparing for a hypothesis |

Step 7
A3: Applied Hypothesis Generator

| formulate new scientific hypotheses from synthesized connections and gaps | Use templates: "If A and B are incompatible but work, latent C is likely" "If X is not explained by Y, but is related to Z, there is probably a generalizing model W" Formulate 2–5 hypotheses at the intersections of models, conflicts, and gaps. | Generative templates GPT systems with instructions (e.g.: " generate hypotheses from conceptual conflicts ") | New hypotheses Architectural drawing of theoretical expansion |

## *3.6. Applying the PANDAVA Protocol: The Case of Abiogenesis*

Let us consider a concrete application of the PANDAVA protocol to the analysis of concepts in abiogenesis — one of the most multidisciplinary and pressing directions in science.

*Step 1. P1: Problem Cartography*

Initial problem: What are the probable scenarios for the emergence of organic molecules capable of leading to life, considering processes on Earth (lightning, alkaline vents, atmospheric turbulence) and in the universe (methanol in trans-Neptunian objects)? Here, we note a new framing of the problem: from the paradigm "Earth and Life" to the paradigm "Cosmos, Proto-Chemistry, Earth, Life."

*Step 2. A1: Argument harvesting*

| **Concept** | **Brief Description** |
|---|---|
| Alkaline hydrothermal vents (Russell & Hall, 1997) | Local self-organization of organic molecules |
| Microlightning in water droplets (Meng, 2025) | Local radical reactions and polymerization |
| Turbiogenesis (Knar, 2025) | Atmospheric vortices and lightning trigger chemical activity |
| Coacervate droplets | Self-assembly in aqueous environments |
| Methanol in TNOs (Brunetto, 2025) | Ancient reservoirs of organics in the Solar System |

Here, we identify five key concepts, each covering different stages and environments.

*Step 3. N1: Navigation through Structures*

We visualize the concept graph of abiogenesis via the PANDAVA model (Figure 7).

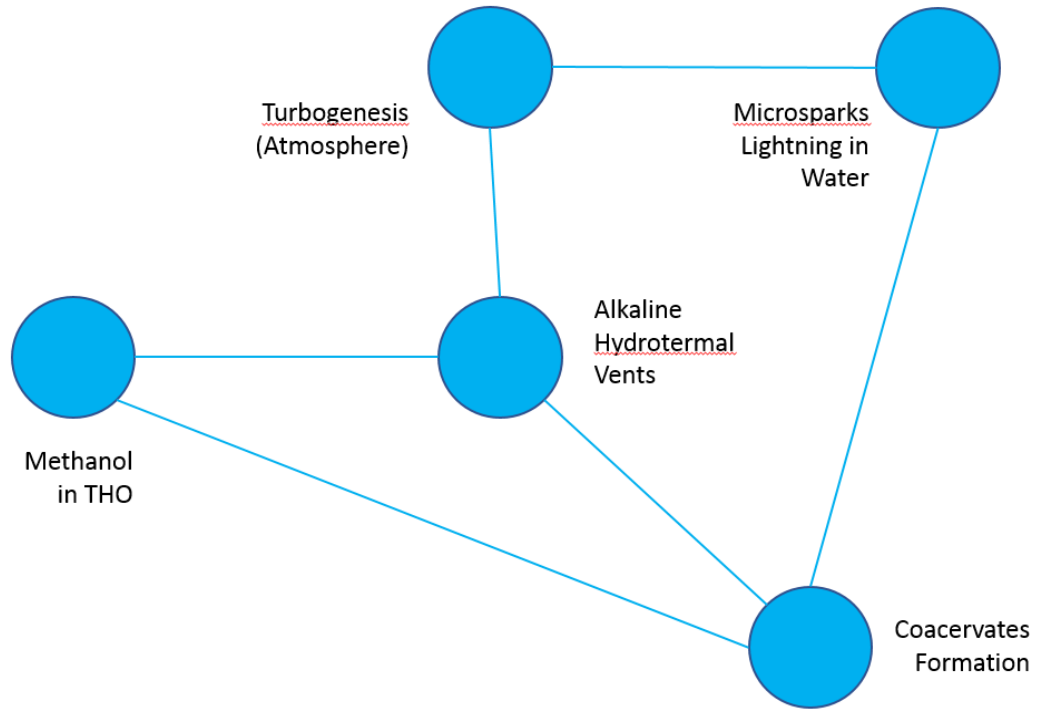

**Figure 7.** Conceptual Graph of Abiogenesis

The following connections are highlighted in the graph:

- methanol in TNOs influences alkaline hydrothermal solvent and coacervate formation (as a source of ancient organic matter),

- Turbulogenesis leads to the formation of microsparks, which in turn contribute to coacervate formation.

-Alkaline hydrothermal solvents also directly facilitate coacervate formation.

In the heatmap (Figure 8), the more saturated the color is, the greater the concept's maturity on that specific criterion. From the map, we can identify the strengths and weaknesses of each concept across all five dimensions.

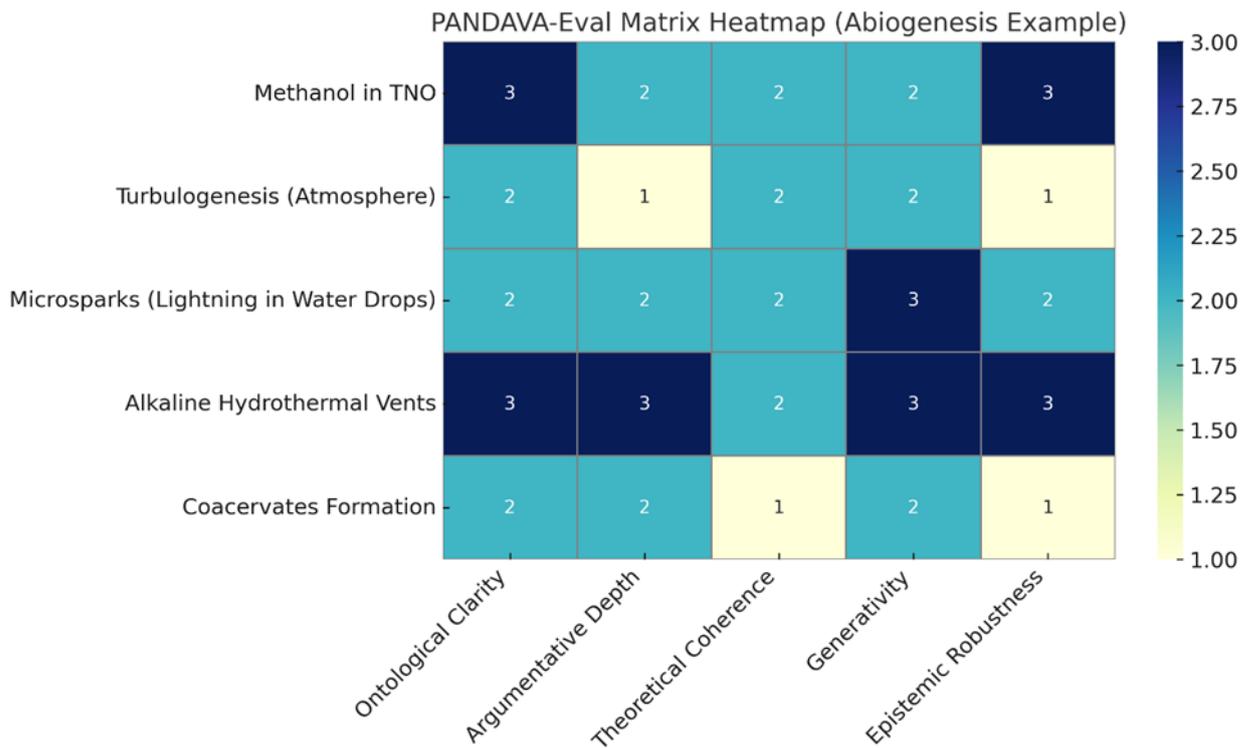

**Figure 8**. Concept Maturity Heatmap (Abiogenesis Example)

Thus, we obtain a structured network of processes that link cosmic, atmospheric, and geochemical sources, with coacervates as the central structural assembly.

*Step 4. D1: Deep Clustering of Concept Maturity*
Maturity scores (PANDAVA-Eval matrix)

| Concept | Ontology | Argumentation | Coherence | Generativity | Robustness |
|---|---|---|---|---|---|
| Alkaline Hydrothermal Vents | 3 | 3 | 2 | 3 | 3 |
| Microsparks (Microlightning) | 2 | 2 | 2 | 3 | 2 |
| Turbulogenesis | 2 | 1 | 2 | 2 | 1 |
| Coacervates Formation | 2 | 2 | 1 | 2 | 1 |
| Methanol in TNOs | 3 | 2 | 2 | 2 | 3 |

Evaluation criteria (5 key dimensions)

| Criterion | What is measured | Questions for evaluation |
|---|---|---|
| Ontological Clarity | How clearly the concept is defined in terms of entities and relationships | Does the concept have a clear ontological structure? |
| Argumentative Depth | How well-substantiated the concept is | Are there rigorous justifications, not slogans? |

| Theoretical Coherence | Integration into existing theories | Does it fit into a coherent explanatory framework? |
| Generativity | Capacity to generate new ideas/connections | Can the concept evolve and expand? |
| Epistemic Robustness | Resistance to critique and alternatives | Can the concept withstand analytical pressure? |

Scoring Scale (0–3)

| Score | Interpretation |
|---|---|
| 0 | No signs of the criterion |
| 1 | Weak or vague presence |
| 2 | Moderate manifestation, partial maturity |
| 3 | Strong manifestation, high maturity |

Detailed Justifications by Concept:

Methanol in TNOs
- Ontological Clarity (3): Clearly defined object and process — methanol in ice layers,
- Argumentative Depth (2): Supported by scientific data, although indirect (spectral interpretation),
- Theoretical Coherence (2): Fits into models of organic origin but lacks direct linkage to prebiotic evolution on Earth;
- Generativity (2): The potential for new models, although still underdeveloped,
- Epistemic Robustness (3): Confirmed by astronomical observations from multiple missions.

Turbulogenesis (Atmosphere)
- Ontological Clarity (2): Concept forming (turbulent vortices), but not strictly defined in the literature.
- Argumentative Depth (1): Logical but lacking direct experimental data.
- Theoretical Coherence (2): Compatible with known atmospheric physics.
- Generativity (2): Generativity can generate multiple chemical activation scenarios.
- Epistemic Robustness (1): Theoretical assumption needs empirical validation.

Microsparks (Lightning in Water Drops)
- Ontological Clarity (2): Clear object - microdischarges in water droplets,
- Argumentative Depth (2): Backed by laboratory experiments,

- Theoretical Coherence (2): Integrated into plasma and hydrodynamic physics,
- Generativity (3): Explains localized chemical reactions,
- Epistemic robustness (2): This is based on real measurements, although under specific conditions.

Alkaline hydrothermal vents
- Ontological Clarity (3): Clearly, described objects - alkaline vents,
- Argumentative Depth (3): Strong experimental support,
- Theoretical Coherence (2): Compatible with chemical evolution models, but not all processes are fully known;
- Generativity (3): Capable of explaining autocatalytic networks, membrane-like structures,
- Epistemic robustness (3): Supported by various independent studies.

Coacervate Formation
- Ontological Clarity (2): The idea of coacervates is understandable but allows various interpretations;
- Argumentative Depth (2): Some experimental basis,
- Theoretical coherence (1): Limited integration with real chemical evolution.
- Generativity (2): high structural variability,
- Epistemic robustness (1): Weak evidence for the transition to fully formed cells.

Interpretation of Maturity Levels
*Most mature concepts:*
Methanol in TNOs and alkaline hydrothermal vents—high ontology, strong argumentation, epistemically robust,
*Moderately mature:*
Microsparks — especially strong in generativity.
*Less mature/fragile zones:*
Turbulogenesis and coacervation—low epistemic robustness and weaker coherence.

Thus, methanol (TNO) and alkaline vapours are mature core concepts. Turbulogenesis and microsparks are related processes that require integration, and coacervates represent secondary assemblies depending on prior chemical mechanisms.

*Step 5. A2: Gap Mapping*
We identify the main gaps in the field of abiogenesis
- No clear model for how methanol delivery is related to Earth-based processes,

- Insufficient integration between atmospheric chemistry and oceanic geochemistry,
- Weakly developed transitions from polymers to membrane structures.

In the radar chart (Figure 9), the closer the line is to the outer edge ("High"), the greater the gaps of the concept on that criterion; the closer to the center ("Low"), the more robust the concept is on that dimension.

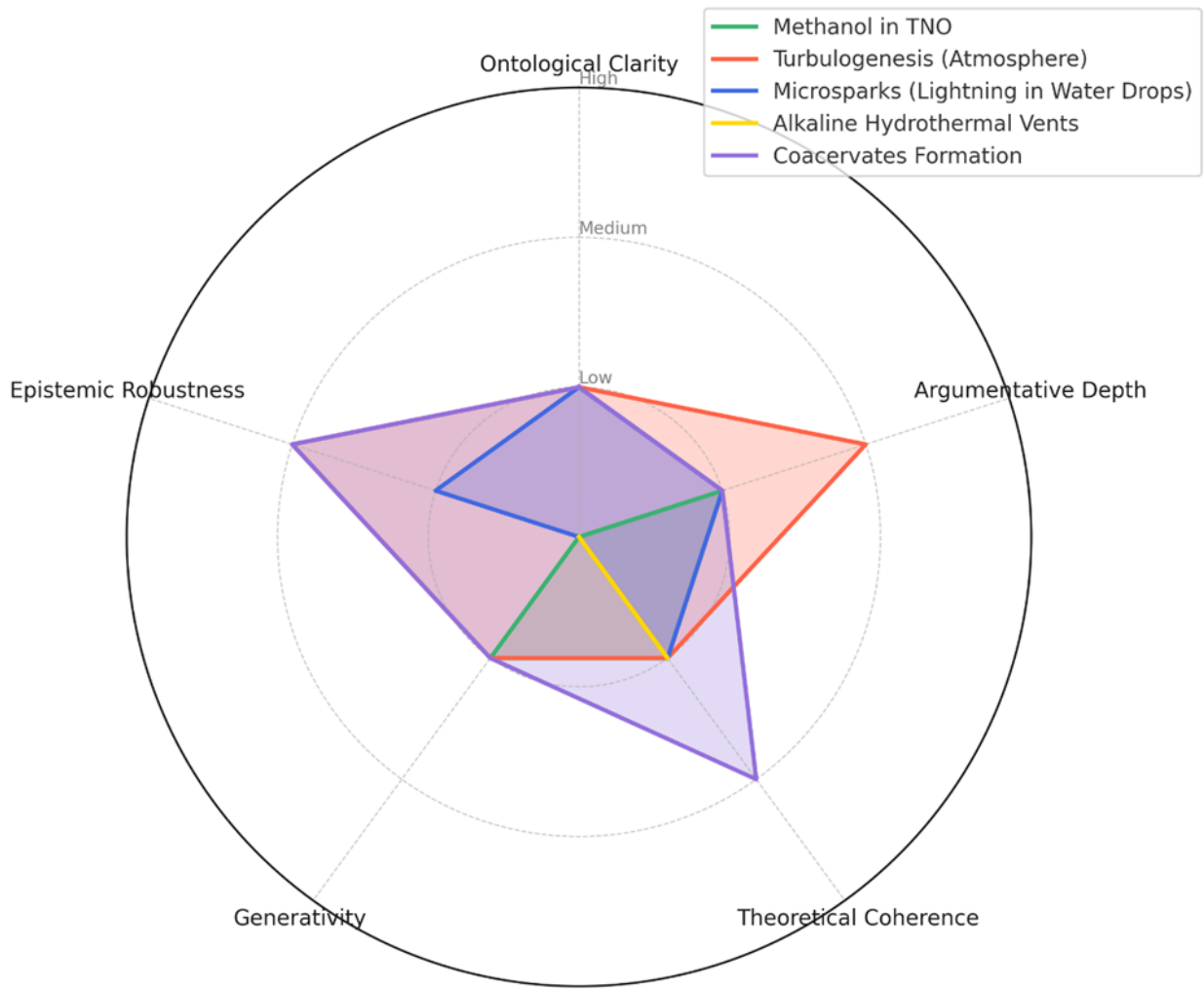

**Figure 9.** Concept Vulnerability Radar (Abiogenesis Example)

*Step 6. V1: Value-oriented synthesis*
We proceed to merge the models:
1. Cosmos:
   Trans-Neptunian Objects → delivery of methanol → formation of a prebiotic organic set.
2. Earth's Atmosphere:
   Turbulent flows + microsparks → generation of active radicals, polymer synthesis.

3. Geochemical structures:
   Alkaline hydrothermal vents + incoming organics → stabilization, catalytic assembly.
4. Structural Organization:
   Coacervates/bubbles as the beginning of spatial compartmentalization.

*Step 7. A3: Applied Hypothesis Generator*
We generate new hypotheses:
1. Multigenesis:
   Life can originate simultaneously in multiple environments, including atmospheric, aqueous, and mineral environments.
2. Methanol-Mediated Evolution:
   Ancient methanol from TNOs acted as a key catalyst for early polymerization.
3. Turbulent-Electrical Trigger:
   Turbulent atmospheric cells accelerated the emergence of molecular complexity even before they contacted liquid water.

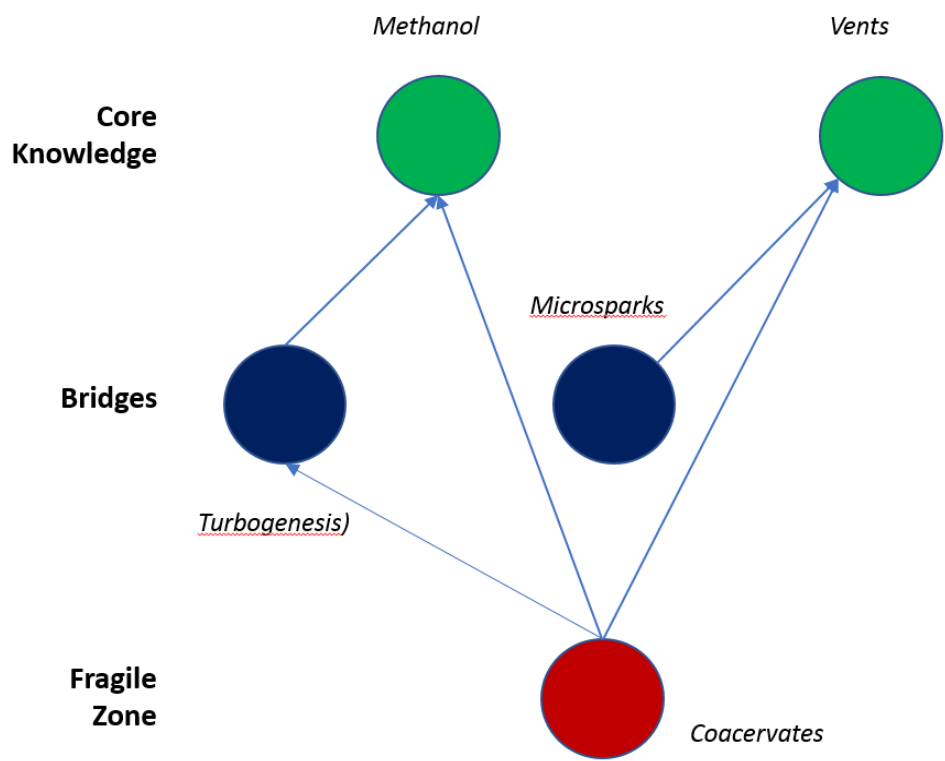

**Figure 10**. Final Architectural Knowledge Synthesis Map (Abiogenesis Example, PANDAVA Protocol)

In this map,

- Core knowledge: *Methanol* and *Vents* constitute the mature core of knowledge.
- Bridges: *Microsparks* and *turbulogenesis* as supporting processes.
- Fragile Zones: *Coacervates* as the domain of vulnerabilities and semantic gaps.

The map represents a multilevel architecture of concepts organized according to maturity and role in the origin-of-life process. The arrows indicate the directions of semantic synthesis between the levels.

- From bridges to the core (supporting basic processes),
- From the core to fragile zones (strengthening weaker areas).

*Core Knowledge (Knowledge Core)*

Methanol (TNO):

Methanol from trans-Neptunian objects — a source of ancient organics. Confirmed by astronomical data (James Webb Telescope).

Alkaline Hydrothermal Vents

Sites of stable autocatalytic chemistry on early Earth.

This approach is extensively supported by experimental evidence (Martin, Russell, et al.).

These concepts form a reliable foundation upon which origin-of-life scenarios can be reconstructed.

*Bridges (Supporting Concepts)*

Microsparks (Lightning in Water Drops):

Local radical reactions triggered by microsparks in water droplets. Capable of generating amino acids and polymers in microenvironments.

Turbulogenesis (Atmosphere):

Atmospheric turbulent vortices create conditions for electrical discharges and localized chemical reactions.

These processes enrich and sustain basic scenarios, providing additional pathways for organic molecule generation.

*Fragile zones (vulnerable areas)*

Coacervate Formation:

Self-assembly of coacervate droplets as prototypes of cellular membranes. Although experimentally confirmed, it has weak theoretical coherence with a complete biogenesis program.

Further support and clarification regarding the transition from random coacervates to organized cellular structures is needed.

*Interpretation of the Synthesis Structures*
1. Core – Bridges:
   Bridges (microsparks, turbulogenesis) enhance the base processes at Vents and around methanol, providing energy and chemical conditions.
2. Core – Fragility:
   Methanol and Vents create conditions for the emergence of Coacervates, but without them, coacervates are unlikely.
3. Need for further reinforcement
   Future research should focus on transitions from unstructured droplets to stable membrane systems and from random polymerizations to autocatalytic reaction networks.

Thus, this architectural model reflects the paradigm that life could not have originated "in a single place." Rather, it was a network of interconnected processes, where strong mechanisms supported and developed weaker links. This viewpoint gives rise to a true architectural understanding of abiogenesis rather than just a list of hypotheses.

Summary of Insights

| PANDAVA Advantage | Explanation |
|---|---|
| Identification of strong and weak zones | Research efforts can be prioritized |
| Understanding integration pathways for concepts | A coherent hypothetical framework is built |
| Opportunity to plan new hypotheses | Targeted research directions |
| Cognitive reconstruction of scientific fields | Dynamic understanding instead of static listing |

Thus, PANDAVA allowed structuring the scattered hypotheses of abiogenesis.

Instead of a mere list (methanol, hydrothermal vents, microsparks, coacervates), an architecture of interconnections was constructed. Core, Bridges, and Fragile Zones within the system of concepts were clearly identified.

Key semantic gaps were also detected, primarily the transition from chemical activity to stable structures and the insufficient integration of cosmic material (methanol) with Earth-based processes.

In general, a rigorous evaluation of concept maturity (PANDAVA-Eval matrix) was conducted. It was shown which ideas are truly robust (*alkaline vents*, *methanol*) and where additional theoretical and empirical work is needed (*coacervates*, *turbulogenesis*).

Additionally, new hypotheses were generated
- Electroinduced mineralized autocatalysis,
- Multigenesis across environments (atmospheric, aqueous, mineral),

- Methanol-mediated early evolution.

Overall Conclusion

Applying the PANDAVA protocol to the domain of the origin of life demonstrated its analytical, evaluative, and methodological value.

| Without PANDAVA | With PANDAVA |
|---|---|
| Separate unrelated hypotheses | Architectural map of interconnections |
| Hidden semantic gaps | Identification of vulnerabilities and growth points |
| Undefined strengths and weaknesses | Objective evaluation of concept maturity |
| Static listing of models | Dynamic knowledge design |

PANDAVA does not merely systematize information. It helps scientists understand where true science is born, where gaps are hidden and where new hypotheses can be developed.

Thus, PANDAVA transforms a systematic review into a cognitive reconstruction of scientific space, paving the way for new discoveries.

Summary: Value of PANDAVA for Science:

| What PANDAVA Provides | Why It Matters for Scientists |
|---|---|
| Structuring of knowledge | Easier to see the holistic picture of the research field |
| Identification of gaps and weak zones | Ability to plan new investigations |
| Evaluation of concept maturity | Prioritization of efforts and resources |
| Synthesis of new hypotheses | Development of paradigms instead of stagnation |
| Architectural thinking | Elevation of research to a higher cognitive level |
| Flexibility across fields (biology, physics, AI, etc.) | Universal applicability across scientific domains |

Thus, PANDAVA is interpreted as a next-generation protocol for scientific thinking, combining strict systematization, reflective diagnostics of knowledge fields, and targeted synthesis of new research. The case of abiogenesis shows that without such a tool, we would only see scattered fragments.

With PANDAVA, we see the scientific field as a living, evolving structure.

## 4. Discussion

Let us discuss the semantic architecture of knowledge within the PANDAVA protocol. Modern systematic reviews typically reduce knowledge to a formal list of articles, stripped of any deep conceptual structure. In contrast, the PANDAVA protocol reveals the hidden architecture of knowledge, represented as an ontologically organized field; a network of arguments, contradictions, and deficits; and a dynamic landscape of scientific maturity and productivity.

Within the context of this new knowledge protocol, we do not simply structure sources—we analyse how knowledge unfolds, where it is fragmented, where synthesis points are hidden, and how new conceptual trajectories are formed from them.

The visualization of the 3D epistemic gap landscape demonstrates that knowledge is neither uniform nor linear.

Elevations (e.g., *ethical AI*) point to inflationary or underdefined concepts.

Plateaus (e.g., *GNN*, *HPO*) indicate the structural cores of scientific maturity.

Ridges and crests (e.g., *Critique of XAI*) represent areas of epistemic friction, where reflection and paradigm struggles occur.

Such a landscape allows for orientation within the cognitive space, avoiding pitfalls—that is, synthesis on the basis of empty concepts—and building meaningful routes through knowledge.

The multilevel synthesis map revealed a clear stratification of concepts

| Level | Function |
| --- | --- |
| Foundational Concepts | Structural foundation for theoretical synthesis |
| Bridge Concepts | Transitional nodes between practice and theory |
| Fragile Concepts | Indicators of cognitive deficits |

This vertical structure allows for conscious hypothesis generation (A3) from mature foundations, the use of intermediate elements as intellectual bridges, and epistemic filtering at the A2 (gap mapping) stage.

The cluster map of the gaps revealed that mature and immature concepts naturally grouped together. It is possible to distinguish
- cluster of conceptual reliability (*GNN + HPO*),
- cluster of critical tension (*XAI*),
- cluster of declarative pseudoideas (*Ethical AI*).

This enables the construction of synthetic trajectories only within or between adequately compatible clusters and the management of concept quality not only at the level of "include/exclude" but also at the level of cognitive compatibility configuration.

Thus, PANDAVA allows us to shift:
- from the traditional logic of

"*We selected 80 articles from database X and then performed aggregation.*"
- to a new, meaning-driven logic of

"*We reconstructed the epistemic map of the field, identified zones of maturity, conflict, and synthesis potential, and then formulated new hypotheses.*"

In our view, this constitutes next-generation metascience, in which:

- Review becomes a form of thinking,
- Visualization becomes a form of logic,
- Synthesis becomes a form of scientific design.

The semantic architecture of knowledge within PANDAVA is not merely a map of concepts; it is a dynamic cognitive system, a tool for intellectual management of knowledge and a means of designing new scientific meaning.

Thus, the protocol not only allows analysis of the existing field but also enables the design of future knowledge, taking into account its logical structure, theoretical tensions, and semantic directionality.

From the above, we can conclude that PANDAVA offers a new position in the philosophical problem (Table 16), replacing binary judgments (agent/nonagent) with architectural synthesis.

**Table 16.** *General conclusions: Why PANDAVA is needed*

| Goal | How It Is Addressed by PANDAVA |
|---|---|
| Find new hypotheses | Through identification of hidden gaps and model comparison |
| Conduct an unconventional systematic review | Through ontology, visualization, and semantic field structuring |
| Justify the conceptual core of an article | Via clustering of mature and productive concepts |
| Build a literature map | Ontological and epistemic mapping + Gap Radar |
| Avoid pseudoscientific inclusions | Through the PANDAVA-Eval Matrix and filtering |
| Stand out in the scientific community | Through semantic depth and intelligence-oriented methodology |

Let us additionally consider the limitations of the PANDAVA protocol in the context of reflecting on the method's application boundaries.

Despite the broad capabilities and potential of the conceptual PANDAVA protocol,

its application is associated with a number of methodological, practical, and cognitive limitations that must be considered when interpreting results. Unlike statistical meta-analyses, PANDAVA operates on conceptual and logical-semantic structures, which implies the following:

- the necessity of manual or semiautomated annotation of theoretical constructs,
- The inclusion of researcher judgment when assigning concepts to clusters or maturity levels.

These circumstances drive the method toward productive reflexivity but also require a high epistemic culture on the part of the user.

The PANDAVA-Eval matrix and gap mapping are based on qualitative evaluations of concepts and rely on maturity, coherence, and productivity scales. However, these scales are not universally standardized, and evaluations may vary among different research groups.

Thus, the protocol does not claim numerical objectivity but offers a cognitive-interpretative tool.

Accordingly, its reliability depends on the transparency of logic and the reasonableness of criteria.

PANDAVA supports integration with NLP systems (BERT, SciSpacy, and GPT). However, automatic extraction does not always correctly recognize ontological levels (theory vs. hypothesis), works poorly with philosophical or interdisciplinary texts and requires manual validation and retraining.

Nevertheless, this limitation can be overcome by a hybrid approach—human–machine collaboration—which is now clearly a major trend.

Moreover, PANDAVA does not replace statistical protocols (PRISMA, MOOSE, and GRADE) when working with clinical trials, experimental meta-analyses, and quantitative evidence integration. Instead, it complements them, allowing meaningful management of concepts, theories, and semantic frameworks.

In general, full application of the protocol requires the following:
- Mastery of semantic and ontological thinking,
- Time for visualization and interpretation,
- Familiarity with the epistemology and philosophy of science principles.

This makes the protocol less accessible for broad use but more valuable for research groups aiming for conceptual leadership.

Thus, the limitations of PANDAVA stem not from deficiencies but from the very nature of the approach: meaning-centered, architectural, and reflexive.

Unlike linear meta-analyses, it requires hypothesis-level thinking, but in return, it provides strategic clarity, structural depth and intellectual meaningfulness of synthesis.

Therefore, these limitations also open development prospects for the protocol, including the automation of concept maturity evaluation, creation of an online interface (PANDAVA-App) and training teams in semantic analysis.

## 5. Conclusion

In conclusion, we reiterate and summarize the statements and results presented in the form of a Technical Protocol for PANDAVA.

Technical Protocol for PANDAVA Application and Simulation

The objective is to develop, implement, and quantitatively verify a novel cognitive–semantic protocol for systematic knowledge analysis oriented toward
- ontological structuring of scientific domains,
- identification of epistemic gaps,
- and synthesis of new hypotheses through semantic mapping.

**Table 17.** Stages of the PANDAVA Protocol

| Stage | Module | Purpose | Output |
|---|---|---|---|
| 1 | P1: Problem Cartography | Ontological mapping of the research problem | Map of entities and levels of analysis |
| 2 | A1: Argument Harvesting | Semantic extraction of theories, models, arguments | Concept catalogue |
| 3 | N1: Navigation through Structures | Building a network of connections between concepts | Conceptual graph |
| 4 | D1: Deep Clustering | Clustering concepts by maturity and coherence | Epistemic clusters |
| 5 | A2: Gap Mapping | Identifying semantic gaps and risk zones | Gap Heatmap, 3D Epistemic Landscape |
| 6 | V1: Value-Oriented Synthesis | Scenario-based synthesis of core and bridge concepts | Architectural knowledge map |
| 7 | A3: Applied Hypothesis Generator | Formulating new hypotheses based on knowledge structure | List of hypotheses and research directions |

PANDAVA application checklist

**Table 18.** PANDAVA application checklist

| № | Stage/Action | Description | Status |
|---|---|---|---|
| 1 | Definition of the research problem | Ontological decomposition of the subject area: entities, processes, levels of analysis | |
| 2 | Formation of the literature corpus | Relevant publications covering key concepts selected | |
| 3 | Semantic annotation of texts | Theoretical models, hypotheses, arguments extracted (manually or via NLP) | |
| 4 | Building a network of semantic links | Conceptual graph of relationships created (e.g., explanation, support, conflict) | |
| 5 | Concept maturity assessment (PANDAVA-Eval Matrix) | Concepts evaluated across five criteria: ontology, argumentation, coherence, generativity, robustness | |
| 6 | Concept clustering | Concepts grouped into core, transitional, and problematic (via KMeans/manual classification) | |
| 7 | Identification of semantic gaps (Gap Mapping) | Gap heatmap and 3D epistemic landscape constructed | |

| 8 | Epistemic landscape analysis | Core knowledge, bridge, and risk zones identified |
| 9 | Value-oriented synthesis of knowledge | Architectural map integrating core and bridge concepts built |
| 10 | Hypothesis generation | New hypotheses formulated at intersections of gaps and mature concepts |
| 11 | Documentation of the process | All protocol stages documented to ensure reproducibility |
| 12 | Critical reflection | Evaluation of strengths and weaknesses of the resulting knowledge map conducted |
| 13 | Final report preparation | Summary materials prepared: maps, diagrams, hypothesis list |

In general, the PANDAVA technical protocol represents a reproducible, cognitively rich, and methodologically formalized system for diagnosing the structure of scientific knowledge, identifying epistemic gaps, and stimulating hypothesis generation and paradigm synthesis.

In this context, the protocol integrates best practices from traditional systematic reviews (e.g., PRISMA) with novel cognitive–epistemic methods.

In this work, we presented and tested a new protocol for systematic conceptual review: PANDAVA (Protocol for Analysis and Navigation of Deep Argumentative and Valued Knowledge).

Unlike existing protocols (e.g., PRISMA, GRADE, and STROBE), which focus on the formal aggregation of empirical data, PANDAVA implements a fundamentally different approach: epistemic reconstruction of the knowledge field on the basis of conceptual cartography, ontological analysis, semantic clustering and the generation of new scientific hypotheses.

The protocol relies on a seven-level structure that includes the modules specified above. This internal structure enables not only systematization but also deep reflection on the state of scientific knowledge—its maturity, internal contradictions, and latent potential for growth.

The protocol's visual instruments (e.g., Gap Radar, Epistemic Landscape, Synthesis Map, and PANDAVA-Eval Matrix) allow researchers to perceive knowledge as a structured space—one in which movement, design, improvement, and connection of ideas is meaningful and intentional.

The application of the protocol to a relevant theory, abiogenesis, demonstrated its universality, adaptability, and high informativeness. PANDAVA not only enables intellectually saturated reviews but also becomes a tool of scientific engineering—that is, a mechanism for creating new theoretical linkages, research trajectories, and cognitive stability.

Thus, PANDAVA sets a new standard for metascientific thinking, one focused not on the sum of publications but on the depth, coherence, and productivity of knowledge as a semantic structure.

The PANDAVA protocol opens new prospects for developing hypothesis-centered science, in which a systematic review becomes an act of constructing the intellectual landscape.

In conclusion, we emphasize that PANDAVA was designed as a cognitive–semantic scenario, but clearly, the protocol requires further development through formalism—combining qualitative and conceptual reasoning with numerical interpretation.